# Multi-output Classification using a Cross-talk Architecture for Compound Fault Diagnosis of Motors in Partially Labeled Condition

Wonjun Yi[a], Wonho Jung[b], Hyeonuk Nam[c], Kangmin Jang[b], Yong-Hwa Park[b]

[a]*Department of Electrical Engineering, KAIST, Daejeon, South Korea*

[b]*Department of Mechanical Engineering, KAIST, Daejeon, South Korea*

[c]*Samsung Electronics, Suwon, South Korea*

**Abstract**

The increasing complexity of rotating machinery and the diversity of operating conditions, such as rotating speed and varying torques, have amplified the challenges in fault diagnosis in scenarios requiring domain adaptation, particularly involving compound faults. This study addresses these challenges by introducing a novel multi-output classification (MOC) framework tailored for domain adaptation in partially labeled target datasets. Unlike conventional multi-class classification (MCC) approaches, the MOC framework classifies the severity levels of compound faults simultaneously. Furthermore, we explore various single-task and multi-task architectures applicable to the MOC formulation-including shared trunk and cross-talk-based designs-for compound fault diagnosis under partially labeled conditions. Based on this investigation, we propose a novel cross-talk architecture, residual neural dimension reductor (RNDR), that enables selective information sharing across diagnostic tasks, effectively enhancing classification performance in compound fault scenarios. In addition, frequency-layer normalization was incorporated to improve domain adaptation performance on motor vibration data.

Compound fault conditions were implemented using a motor-based test setup and evaluated across six domain adaptation scenarios. The experimental results demonstrate its superior macro F1 performance compared to baseline models. We further showed that the structural advantage of RNDR is more pronounced in compound fault settings through a single-fault comparison. We also found that frequency-layer normalization fits the fault diagnosis task better than conventional methods. Lastly, we analyzed the RNDR with various conditions, other models with increased number of parameters, and compared with the ablated RNDR structure.

*Keywords:* Compound fault, Partially-labeled condition, Multi-output classification, Cross-talk, Frequency layer normalization, Residual neural dimension reductor

*Email addresses:* lasscap@kaist.ac.kr (Wonjun Yi), wonho1456@kaist.ac.kr (Wonho Jung), hu.nam@samsung.com (Hyeonuk Nam), kangmin55@kaist.ac.kr (Kangmin Jang), yhpark@kaist.ac.kr (Yong-Hwa Park)

## 1. Introduction

With the widespread adoption of rotating machinery in industrial applications, the development and implementation of advanced fault diagnosis systems have become imperative. These systems are vital for ensuring operational efficiency, minimizing unexpected downtime, and avoiding critical failures [1–3].

Fault classification in rotating machinery faces substantial challenges due to the scarcity of labeled data under diverse operating conditions, as manual annotation is both time-consuming and costly. Moreover, variations in load, speed, and environmental conditions across domains significantly alter the spectral and temporal characteristics of fault signals, making it difficult for pre-trained diagnostic models in a source domain to adapt effectively to a target domain. To address this, research has increasingly focused on domain adaptation scenarios where only a small portion of the target domain data is labeled [4–8].

The occurrence of compound faults, such as the simultaneous occurrence of inner race fault (IRF), outer race fault (ORF), misalignment, and unbalance, introduces additional diagnostic challenges. These faults often result from prolonged operational stress or harsh environments, leading to multiple independent fault mechanisms within the same machinery [2, 3, 9]. Researchers have explored these compound fault scenarios [10–15].

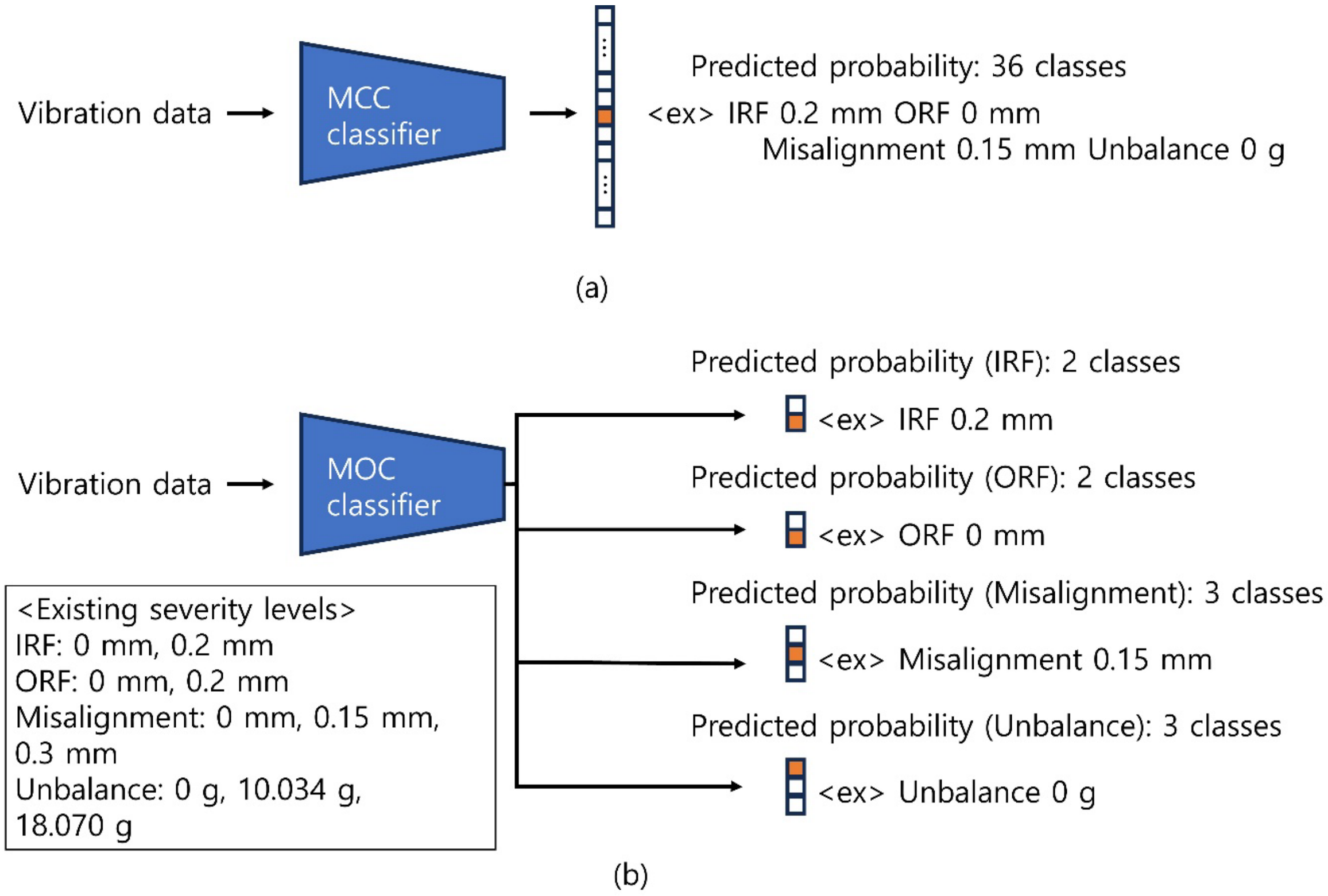

**Fig. 1.** Conceptual comparison between multi-class classification (MCC) and multi-output classification (MOC) approaches for compound fault diagnosis. (a) In the MCC framework, all possible fault combinations are treated as a single classification task with a unified output vector. (b) In contrast, the MOC framework decomposes the problem into multiple sub-tasks, each corresponding to a specific fault type-such as IRF, ORF, misalignment, and unbalance-allowing each to output its own predicted severity level independently.

However, prior research only dealt with domain adaptation on single faults or compound faults without a domain adaptation scenario. So, in our research we incorporate both domain adaptation and motor compound fault classification for problem formulation.

Conventional multi-class classification (MCC) [10, 11] attempts to address these challenges by treating each possible fault combination as a single class, outputting a single predicted probability vector. Since prior research was based on datasets involving only a few distinct fault combination classes, there was little motivation to pursue alternatives to the multi-class classification (MCC) framework. However, this approach becomes inefficient and impractical as the number of fault combinations increases [16]. In particular, some existing studies consider only seven [10] or thirteen [11] compound fault classes, which fall short of representing the combinatorial diversity observed in real-world applications.

A promising alternative to MCC is multi-output classification (MOC), which independently classifies the severity levels of each fault component. Unlike conventional MCC, which regards a combination of faults as a single class like **Fig. 1 (a)**, MOC independently classifies each fault, depicted as **Fig. 1 (b)**. This structural difference in MOC framework reduces class complexity and inter-class interference while simplifying decision boundaries, much effective in classification, especially for domain adaptation scenarios.

In addition to demonstrating the superiority of the MOC framework over traditional MCC, this study explores different approaches to implementing MOC with multi-task learning (MTL) and single-task learning (STL)._This crosstalk mechanism is essential to mimic the fundamental characteristic of a compound fault since a compound fault signal is not simply the sum of each fault component signal, but the fault component interacts with each other in a nonlinear way, making the compound fault signal affected by these interactions [17, 18]. When tasks involve classifying physically related quantities, MTL has been shown to improve classification performance compared to STL [19]. Leveraging this phenomenon, a model learned with MTL that can utilize that information for classifying compound faults can perform better than STL where each classification task of fault component is trained as a separate model.

Moreover, different fault types typically manifest in distinct time-frequency regions, making it difficult for a shared representation to effectively capture all relevant features. Shared trunk-based architectures, such as those in [20, 21], apply a single feature extractor across all tasks. However, this design may lead to conflicting feature usage across tasks, increasing the risk of negative transfer when tasks focus on disparate signal characteristics. To mitigate this, we adopt task-specific feature extractors, each dedicated to a particular fault, and enable selective information exchange through cross-talk layers (CTLs). This design facilitates positive transfer while preserving task-specific focus and has shown improved generalization in multi-task scenarios [22–24]. In the compound fault classification problem, this advantage can make each feature extractor focus on each fault component while learn the relationship between fault components when a compound fault occurs. To further enhance this framework, we propose an improved cross-talk architecture, residual neural dimension reductor (RNDR), that expands upon prior models [22–24] by refining the CTL design to better capture fault interactions and adapt to task-specific requirements. Specifically, RNDR first concatenates information from each convolutional block of feature extractors except the feature extractor of the main task in channel dimension. Then it integrates information with the nonlinear transformation in order to learn the nonlinear relationship between fault components. RNDR block consists of a two-dimensional convolutional neural network (2D-CNN) in order to maintain time-frequency representation of features, enabling the model to learn relationships dependent on frequency components. This architectural enhancement contributes to greater robustness and classification performance, as demonstrated in our experimental results.

In addition, we apply frequency layer normalization (FLN) [25] to reflect the mechanical characteristics of the motor. Originally proposed in our prior work, FLN is designed based on the stationary nature of vibration signals and their dependence on rpm. While not the main contribution of this study, FLN serves as a complementary component that improves feature stability and domain-invariant representation, as confirmed through comparative experiments with conventional normalization methods.

The RNDR is validated using a motor dataset encompassing diverse fault scenarios and operating conditions. Experiments are performed under six domain adaptation cases, and the model performance is evaluated in partially labeled conditions. The macro F1 score is used as the primary performance metric, ensuring fair evaluation across all fault types. During training, the model leverages multi-kernel maximum mean discrepancy (MKMMD) [26] and entropy minimization (EM) [27] as fixed domain adaptation strategies to reduce the distribution gap between the source and target domains.

The remaining paper is structured as follows. Section 2 reviews related work, discussing advancements and challenges in fault diagnosis and domain adaptation. Section 3 outlines the problem formulation and defines key scenarios. Section 4 describes the RNDR framework. Section 5 details the experimental setup, including dataset preparation, model structure, training process, and evaluation protocols. Section 6 presents the results. Section 7 discusses the result and further analyzes the classification performance RNDR architecture under various conditions and ablate structural components of RNDR. Section 8 presents the concluding remarks and highlights directions for future work.

## 2. Related work

### 2.1. *Partially labeled datasets*

Several studies have explored fault classification in partially labeled datasets without incorporating domain adaptation techniques [4–6]. These approaches typically use both labeled and unlabeled data to improve diagnostic accuracy. Methods such as generative adversarial networks [28] for feature extraction [4, 5] and dimensionality reduction techniques [6] leverage structural patterns within unlabeled data, enabling effective classification even under limited labeling conditions.

Alternatively, domain adaptation techniques have been applied to address fault classification in partially labeled datasets. These approaches focus on transferring knowledge from well-labeled source domains to target domains with limited labeled data [7, 8]. For instance, domain adversarial neural networks [29] for aligning source and target distributions [7] and deep adaptive autoencoders combined with manifold learning for enhancing feature extraction [8] effectively leverage structural information within labeled and unlabeled data, leading to improved diagnostic performance under such constraints.

### 2.2. *Compound fault diagnosis system*

Unlike MCC [10, 11], multi-label classification (MLC) frameworks [12, 13] independently assign labels to each fault. MLC can determine the presence or absence of each fault label, but can not capture additional details such as severity level of each fault component.

MOC enables the simultaneous classification of compound faults and their severity levels, and is scalable to compound fault scenarios [14, 15]. However, existing study has classified faults based on both fault type and severity rather than treating each fault separately [14]. This approach lacks generalizability, as fault severity can vary significantly across different fault types. While other studies have attempted to classify fault severity for each fault type [15], they merely demonstrated the feasibility of using an MOC framework without establishing its superiority over MCC. Moreover, none of the existing studies have considered domain adaptation under partially labeled conditions. On the other hand, this paper demonstrates that MOC is more advantageous than MCC for compound fault diagnosis under a partially labeled target domain dataset.

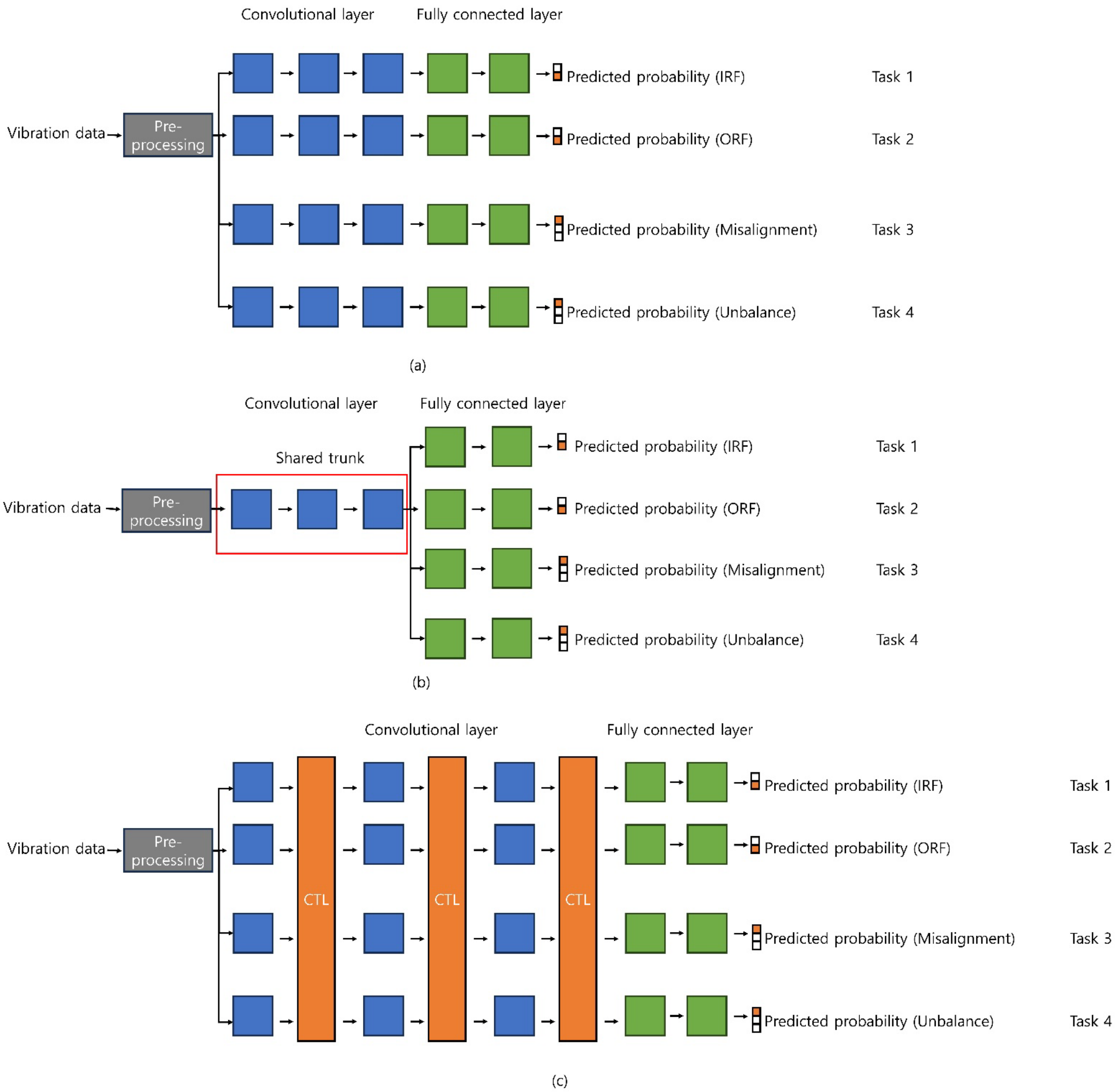

**Fig. 2.** Illustration of the model architectures of MOC. (a) STL: Each task has an independent architecture. (b) Shared trunk: A shared trunk serves as the feature extractor, and the output is processed through task-specific layers corresponding to task 1 (IRF), ..., task 4 (Unbalance). (c) Cross-talk: Convolutional blocks of the feature extractor and task-specific layers are identical to (a). However, feature from the previous convolutional block is fed to the next convolutional block of a different task after fed to the cross-talk layer (CTL).

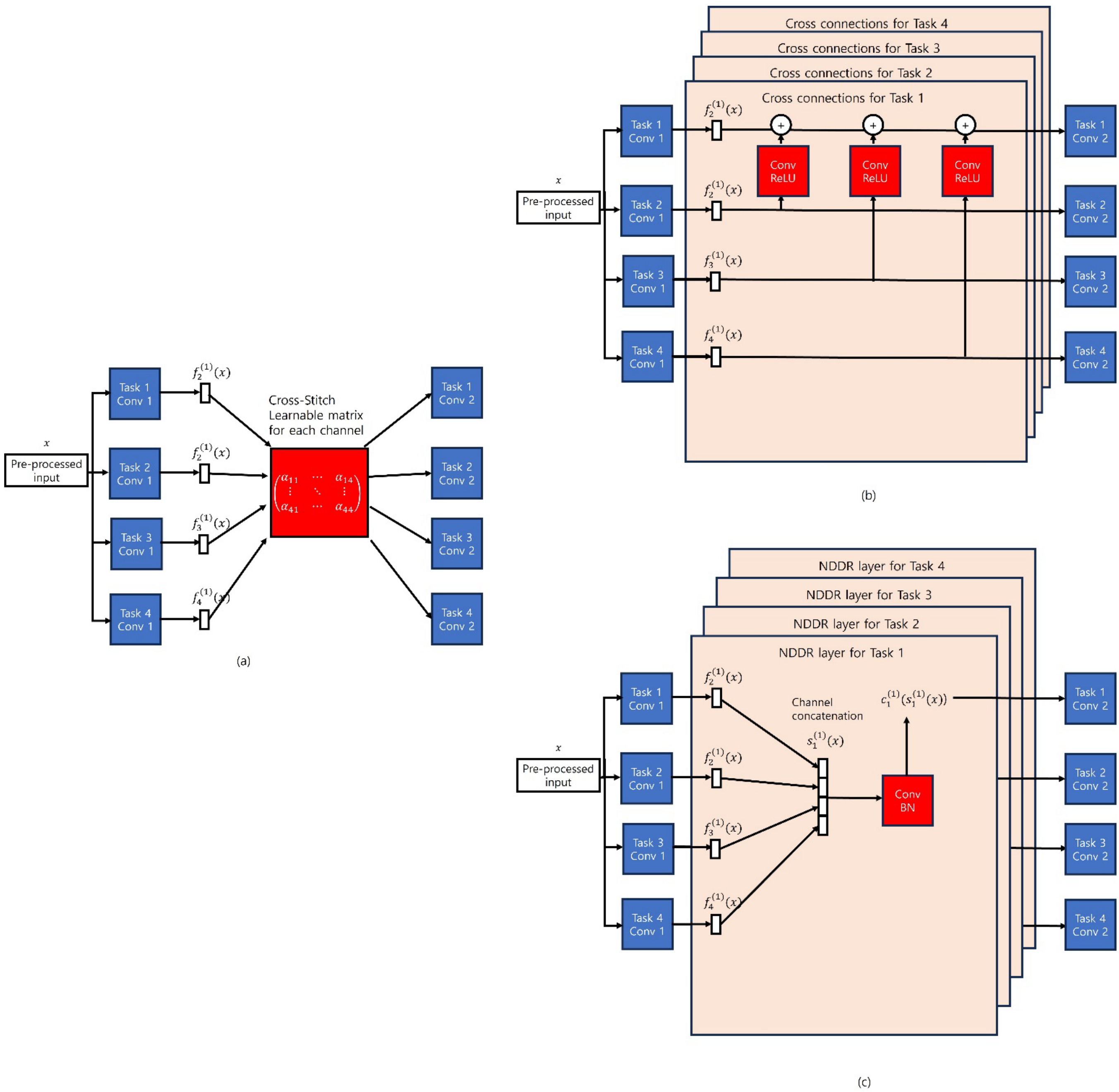

**Fig. 3.** Illustration of the cross-talk architectures. (a) Cross-stitch, (b) Cross-connected, (c) Neural discriminative dimensionality reduction convolutional neural network.

*2.3. Multi-task learning with cross-talk architecture*

The shared trunk approach [20, 21] incorporates a common feature extractor, referred to as the shared trunk, which processes input features before task-specific layers perform individual tasks. This can be shown in **Fig. 2 (b)**. Unlike **Fig. 2 (a)**, a shared trunk is used for all tasks, and fed to each task specific layers (TSL). In the context of MOC, each task corresponds to the severity classification of a specific fault. For instance, task-constrained deep convolutional network (TCDCN) employs a single shared trunk with multiple simple task-specific branches for classification [20], while multi-task attention network (MTAN) shares a shared trunk but additionally applies independent attention blocks for each task [21].

Meanwhile, the cross-talk approach maintains separate feature extractors for each task while allowing information exchange through CTL at each convolutional block [22-24]. **Fig. 2. (c)** shows the concept of cross-talk models. Unlike **Fig. 2 (a)**, cross-talk architecture shares features from convolutional layers through cross-talk layer (CTL). Researchers developed various CTL architectures. For example, as shown in **Fig. 3 (a)**, cross-stitch networks (CS) use learnable matrices for each channel to linearly combine features from different tasks before passing them to the next convolutional block [22]. On the other hands, as **Fig. 3 (b)**, cross-connected networks (CC) extend this idea by using convolutional blocks instead of learnable matrices, enabling channel-specific weight adjustments [23]. Also CC used nonlinear activation, ReLU activation [30]. Neural discriminative dimensionality reduction convolutional neural network (NDDR-CNN) concatenates feature outputs along the channel dimension and applies task-specific NDDR layers before feeding them into the next convolutional block of each feature extractor [24] as **Fig. 3 (c)**. NDDR layer is consisted with convolutional block with kerel size 1, and batch normalization (BN) [31]. Building on these advancements, we developed RNDR.

## 3. Problem definition

In a compound fault classification task, a system may have multiple independent types of faults (e.g. IRF, ORF, unbalance, misalignment), each of which can possess a set of discrete states (e.g. normal, low severity, high severity). Formally, let the $i$-th fault have $m_i$ possible states, represented as the set $D_i$ which can be denoted as following Eq. (1).

$$D_i = \{s_i^1, \dots, s_i^n\} \tag{1}$$

where $s_i^j$ denotes the $j$-th state of $i$-th fault. For a system with $n$ such faults, the entire set of possible system states $D$ can be represented as a Cartesian product of the states of individual faults:

$$D = D_1 \times \dots \times D_n \tag{2}$$

Each state $d \in D$ represents a unique combination of states across the $n$ faults, resulting in $|D| = \prod_{i=1}^{n} m_i$ total combinations of states.

In the MCC approach, each possible combination $d \in D$ is treated as a unique class, resulting in a classification problem with $|D|$ classes. Each class label $y$ is defined as a one-hot vector corresponding to the $d$-th class and can be expressed as:

$$y = e_d, \tag{3}$$

where $e_d$ is a unit vector. On the other hand, in the MOC approach, classification of the severity of each fault is done separately. In other words, the state of each fault $d_i \in D_i$ is classified separately, transforming the original multi-class problem into $n$ parallel classification tasks, each with $m_i$ possible states. Here, each output is expressed as:

$$y_i = e_{d_i}, \tag{4}$$

where $e_{d_i}$ is a unit vector.

The overall system output can be represented as a collection of these independent one-hot vectors:

$$\boldsymbol{y} = [y_1, \dots, y_n]. \tag{5}$$

This MOC approach reduces complexity by decomposing the high-dimensional multi-class problem into multiple lower-dimensional classification tasks.

The partially labeled conditioned target domain dataset contains significantly more unlabeled data than labeled data.

## 4. Proposed method

Under explanations about general cross-talk architecture explained in section 2, the proposed method, RNDR, will be explained in detail, including pre-processing, feature extractors, and TSLs in this section.

### 4.1. *Feature extractor*

Table 1 Feature Extractor Architecture

| **Data flow** | **Module** | **Output shape** |
|---|---|---|
| Input normalization | FLN | $(B, C = 1, F = 80, T = 48)$ |
| | Conv 1 | $(B, C = 32, F = 40, T = 24)$ |
| Convolutional blocks | Conv 2 | $(B, C = 64, F = 20, T = 12)$ |
| | Conv 3 | $(B, C = 128, F = 1, T = 1)$ |

The single-channel vibration data are transformed using a short-time Fourier transform (STFT), denoted as $x \in R^{B\times1\times F\times T}$, where $B$, $F$, and $T$ represent the batch size, frequency, and time, respectively. Each batch included 16 signals. The STFT was performed with a window length of 4096 samples and an overlap of 50%, corresponding to a frame shift of 2048 samples. Then, we cut frequency rage to 20~520 Hz to effectively filter out low and high frequency noise while retaining fault-related features. This results $B = 16, C = 1, F = 80, T = 48$. For the last step of pre-processing, we converted STFT into a dB scale.

Pre-processed input $x$ is then passed through 2D-CNN based feature extractor for each fault. The architecture of feature extractors for each fault is explained in **Table 1**. First, we apply FLN [25] for input normalization before feeding to the convolutional blocks. Then, features are extracted through a series of three convolutional blocks, each consisting of a 2D-CNN, FLN [25], dropout [32], ReLU activation [30], and average pooling operations. These conv blocks progressively extract feature representations, and the final conv block applies an adaptive average pooling operation and removes the time-frequency dimension to reduce the spatial dimensions of the feature maps, resulting in a compact feature vector with dimensions of $(B = 16, C = 128)$.

Here, we applied a FLN [25], specifically designed to normalize vibration signals characterized by dominant frequency components and relative stationarity. Unlike conventional layer normalization (LN) [33], FLN computes the mean and standard deviation along the frequency dimension $F$ for each data sample, channel, and time step independently. The formulation is as follows:

$$Y_{c,:,t} = \gamma \frac{X_{c,:,t} - \mu_F^{(c,t)}}{\sqrt{\sigma_F^{(c,t)^2} + \epsilon}} + \beta, \tag{6}$$

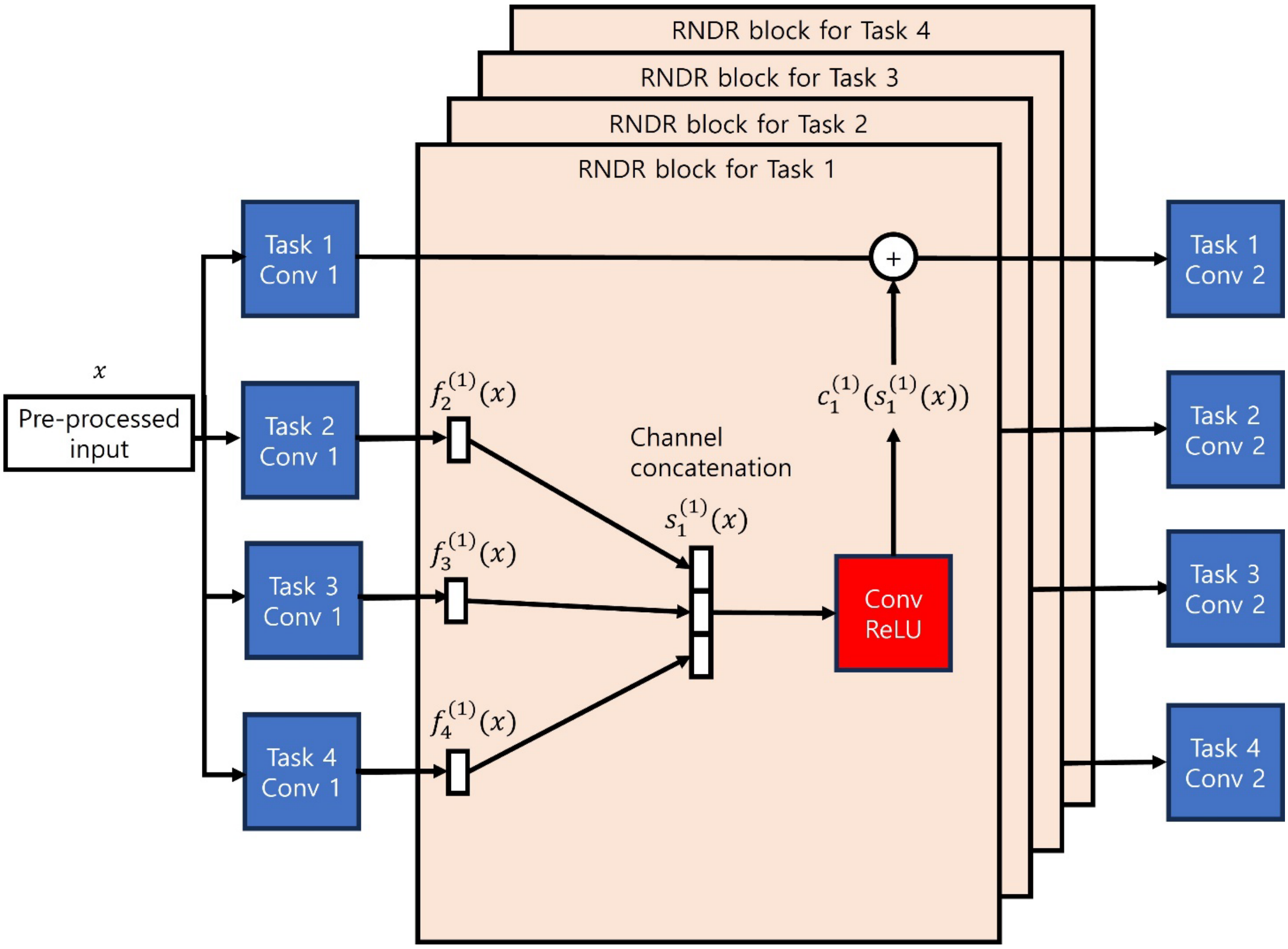

**Fig. 4.** Illustration of the RNDR block between first convolutional layer and second convolutional layer.

where $X \in R^{C\times F\times T}$ is the input variable for a single sample, and $Y \in R^{C\times F\times T}$ is the corresponding output. The mean and variance are computed along the frequency axis $F$ as follows:

$$\mu_F^{(c,t)} = \frac{1}{F}\sum_{f=1}^{F} X_{c,f,t},\ {\sigma_F^{(c,t)}}^2 = \frac{1}{F}\sum_{f=1}^{F}\left(X_{c,f,t} - \mu_F^{(c,t)}\right)^2, \tag{7}$$

where $\mu_F^{(c,t)}$ and $\sigma_F^{(c,t)}$ denote the mean and standard deviation for the frequency dimension $F$ at each channel $c$ and time step $t$. Here, the learnable parameters $\gamma \in R^F$ and $\beta \in R^F$ are applied independently to each frequency component.

### 4.2. *Residual neural dimension reductor block*

The output from a convolutional block in one feature extractor is combined with the output of the RNDR block before being passed to the next convolutional block. RNDR is designed to transfer useful information from other classifiers. Some features captured by convolution blocks of other classifiers can be useful since fault components of compound faults are affecting each other. The computation process of the RNDR block between the classifier for IRF and the others placed on the first convolutional stage is illustrated as an example in **Fig. 4. Fig. 4** displays specifically about RNDR block for task 1 between first and second convolutional layers.

First, for the outputs of first convolutional block of other tasks, $f_2^{(1)}(x)$, $f_3^{(1)}(x)$ and $f_4^{(1)}(x)$ are concatenated along the channel dimension. Here, $f_i^{(j)}(x)$ means output of $j$-th convolutional block for task $i$, and $\odot$ indicates concatenation among the channel dimensions. Then concatenated feature can be denoted as

$$s_1^{(1)}(x) = f_2^{(1)}(x) \odot f_3^{(1)}(x) \odot f_4^{(1)}(x). \tag{8}$$

$s_i^{(j)}(x)$ is concatenated feature of RNDR block between $j$-th convolutional block and $j+1$-th convolutional block for task $i$. This concatenated feature has dimension of $s_1^{(1)}(x) \in R^{B\times 3C\times F\times T}$ when $f_2^{(1)}(x), f_3^{(1)}(x), f_4^{(1)}(x) \in R^{B\times C\times F\times T}$. To ensure proper dimensionality, $s_1^{(1)}(x)$ undergoes a learnable transformation $c_1^{(1)}$ using a 2D-CNN followed by a ReLU activation, restoring it to the original channel size, $c_1^{(1)}(s_1^{(1)}(x)) \in R^{B\times C\times F\times T}$. Like above notation, $c_i^{(j)}$ is notation of nonlinear transformation for dimension reduction of RNDR block between $j$-th convolutional block and $j+1$-th convolutional block for task $i$. During this nonlinear transformation, RNDR can integrate multiple channel information of other classifiers to make useful for classifying fault type while maintaining time-frequency feature. Since, fault components of compound fault are nonlinearly related to each other, nonlinear transformation of multiple channels can make classifier to learn nonlinear relationship. Finally, $c_1^{(1)}(s_1^{(1)}(x))$ is added to $f_1^{(1)}(x)$ and passed as input to convolutional block for next stage. This input can be notated as below.

$$f_1^{(2)}(x) = f_1^{(1)}(x) + c_1^{(1)}(s_1^{(1)}(x)) \tag{9}$$

Here, we added a residual connection to stabilize gradient flow, mitigating vanishing and exploding gradients and easing optimization. In other words, the classifier can exploit information propagated from convolutional blocks of other feature extractors without losing its discriminative focus on its designated fault component.

### 4.3. *Task-specific layer*

Each feature vector from each feature extractor for each fault is processed through TSLs. TSL is composed of the first fully connected layer and the second fully connected layer. The first layer has ReLU activation. In the partially labeled condition, where most target domain data is unlabeled, unsupervised domain adaptation loss (UDA) is necessary to utilize unlabeled data effectively. To achieve this, MKMMD [26] is applied to outputs from penultimate layers, aligning task-specific feature extractors between domains. Each feature vector is then passed through an individual SoftMax classifier, which outputs predicted probability vectors. Classification is optimized using categorical cross-entropy (CCE), with EM [27] further refining predictions by encouraging confident outputs on unlabeled target data.

## 5. Experimental setup

### *5.1. Motor setup and data acquisition*

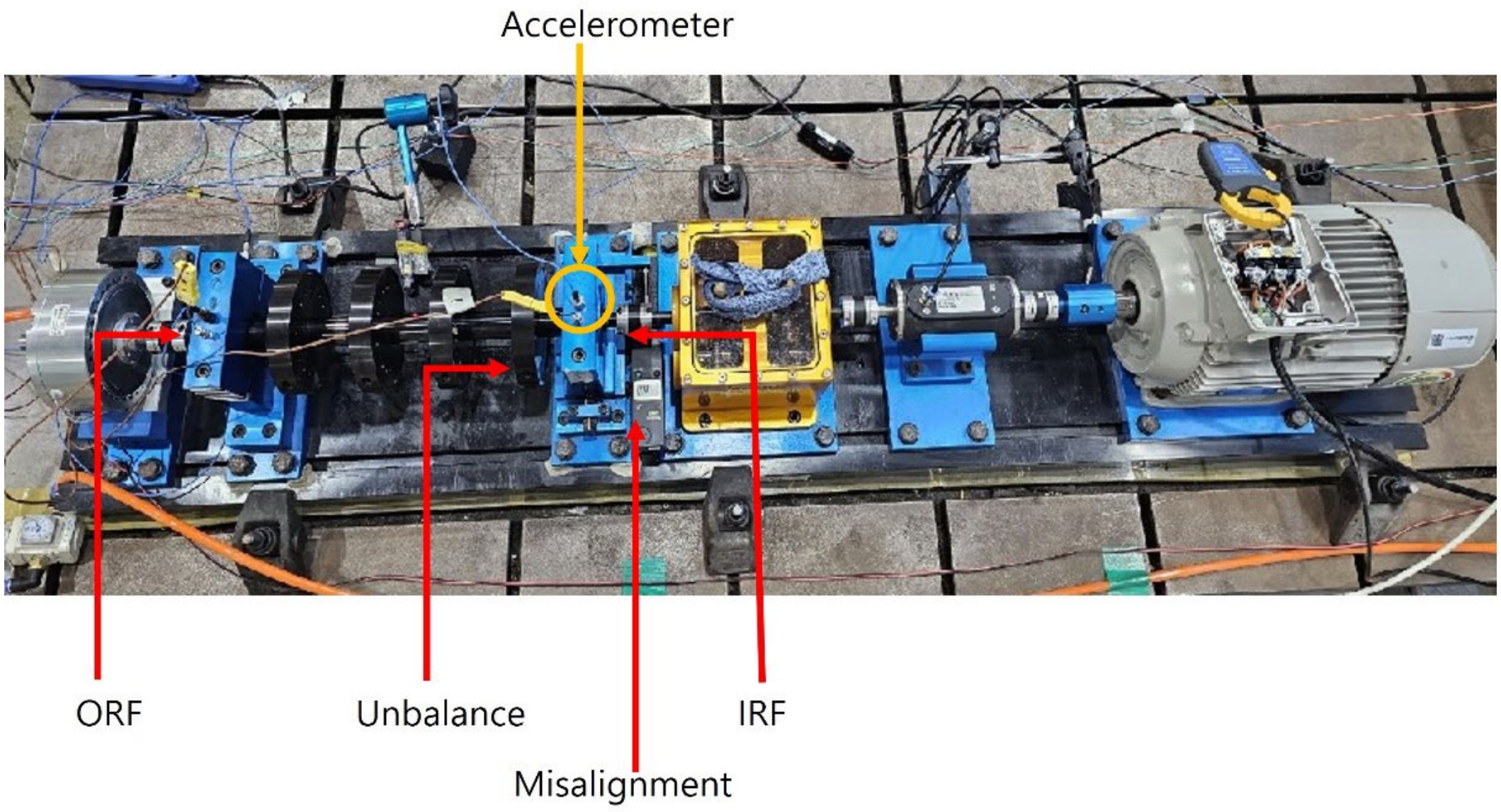

**Fig. 5.** Motor setup for fault diagnosis. Accelerometer was installed vertical to ground, marked as yellow circle. IRF, ORF occur in each bearing, while misalignment and unbalance occur in rotor, marked as red arrow.

We conducted experiments using a motor described in [34], as Fig. 5. The motor setup has the following components (right to left from **Fig. 5**): a motor, a torque meter, a gearbox, a bearing housing used for measurement, a rotor, an additional bearing housing, and a hysteresis brake. The motor was a 3-phase induction motor manufactured by SIEMENS with 3 horsepower and a 4-pole AC configuration. The gearbox had a gear ratio of 2.07. The bearing within the bearing housing was a standardized steel NSK bearing (NSK 6205), with a ball diameter of 7.90 mm, a pitch diameter of 38.5 mm, a contact angle of 0°, and nine balls. Bearing housing is located before and after the rotor. Then, a hysteresis brake (AHB-3A) manufactured by Valid Magnetic Ltd is installed. Torque load is electromagnetically applied to the hysteresis brake. The vibration data was collected using an accelerometer (PCB35234), which is installed on the bearing housing on the right, installed vertical to the ground, shown as orange circle in **Fig. 5**. The accelerometer was installed in accordance with the vibration installation guide [35]. Sampling frequency is set to 25.6 kHz, and measurements are expressed in terms of gravitational acceleration ($9.8\ \mathrm{m}/s^2$). During the experiments, data was recorded as TDMS files using the NI FlexLogger software and subsequently converted to NumPy files. Each dataset was segmented into 4 second intervals.

Table 2 Details of Data Subsets

| **Subset** | Subset A | Subset B | Subset C |
|---|---|---|---|
| **rpm pattern** | Sinusoidal function | Triangular function | 5 Constants |
| **Base rpm ($rpm$)** | 3000 | 4000 | 1800-3000 |
| **Modulation Period ($s$)** | 10 | 5 | N/A |
| **Torque Load ($N \cdot m$)** | Random | 0 | 0 |
| **Record time ($min$)** | Normal: 50<br>Fault (35 types):<br>5 per type<br>Total: 225 | Normal: 50<br>Fault (35 types):<br>5 per type<br>Total: 225 | Normal: 10 per rpm<br>Fault (35 types):<br>1 per type and rpm<br>(5 constants)<br>Total: 225 |

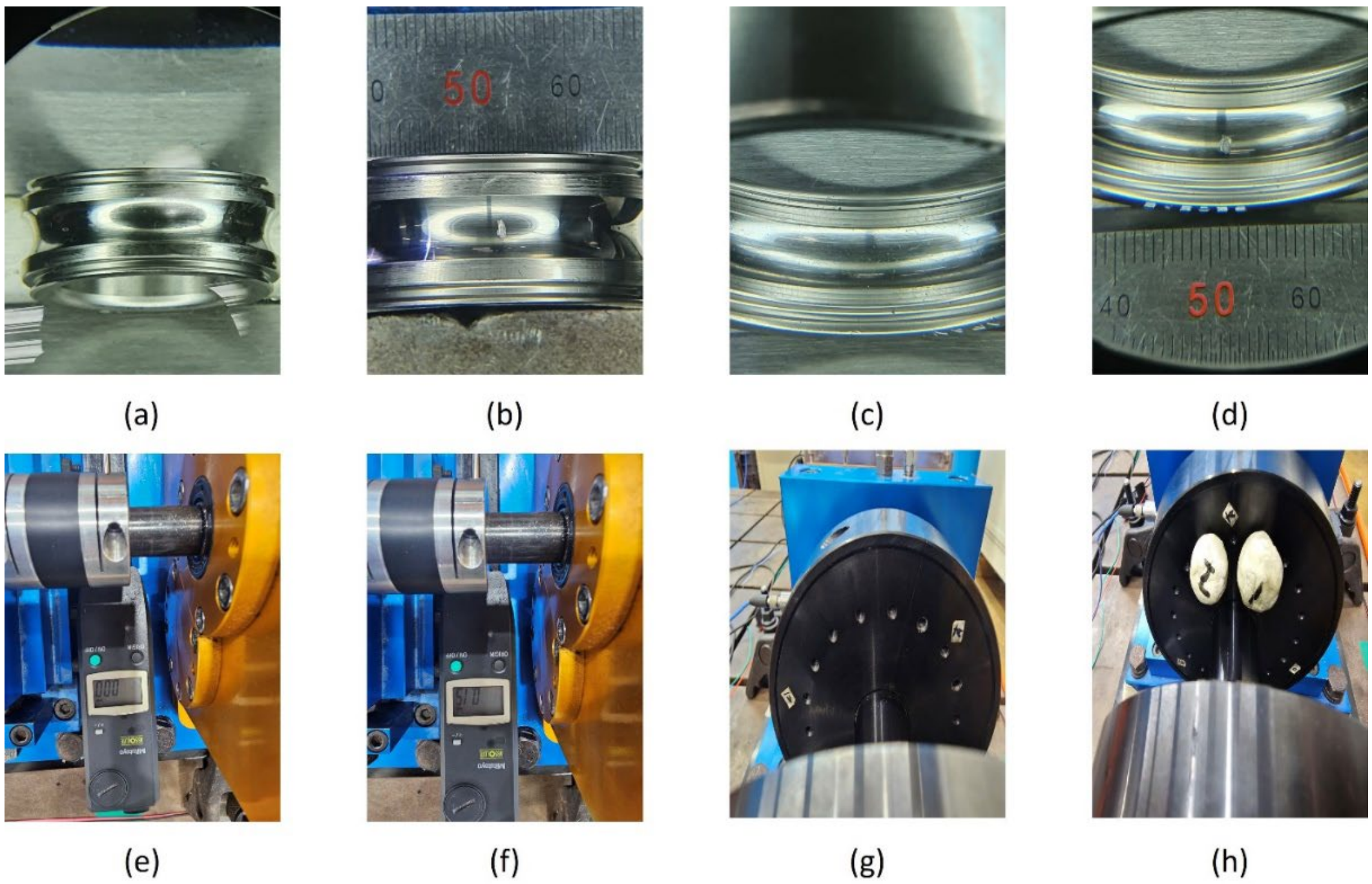

**Fig. 6.** Visual comparison of normal and fault conditions for each fault in the motor's compound fault setup. (a) Inner race normal, (b) inner race fault, (c) outer race normal, (d) outer race fault, (e) normal state without misalignment, (f) fault state with misalignment, (g) normal state without unbalance, and (h) fault state with unbalance.

Under this setup, we applied compound faults consisting of IRF, ORF, misalignment, and unbalance as **Fig. 6.** As pointed out on **Fig. 5**, IRF and ORF occur in bearing, while misalignment and unbalance occur in rotor. In detail, IRF and ORF were applied by creating small artificial cracks in the bearings, as shown in **Fig. 6 (b)** and **Fig. 6 (d)**. For the IRF, two severity levels were defined: normal (0 *mm*) and 0.2 *mm*. Similarly, the ORF had two severity levels: normal (0 *mm*) and 0.2 *mm*. Misalignment was injected by deliberately misaligning the rotor downward based on **Fig. 5**, with three severity levels: normal (0 *mm*), 0.15 *mm*, and 0.3 *mm*. This is portraited in **Fig. 6 (f)**. Unbalance was implemented by attaching masses to the rotor on the far-right side, with three severity levels: normal (0 *g*), 10.034 *g*, and 18.070 *g*. In total, 36 possible fault combinations (2

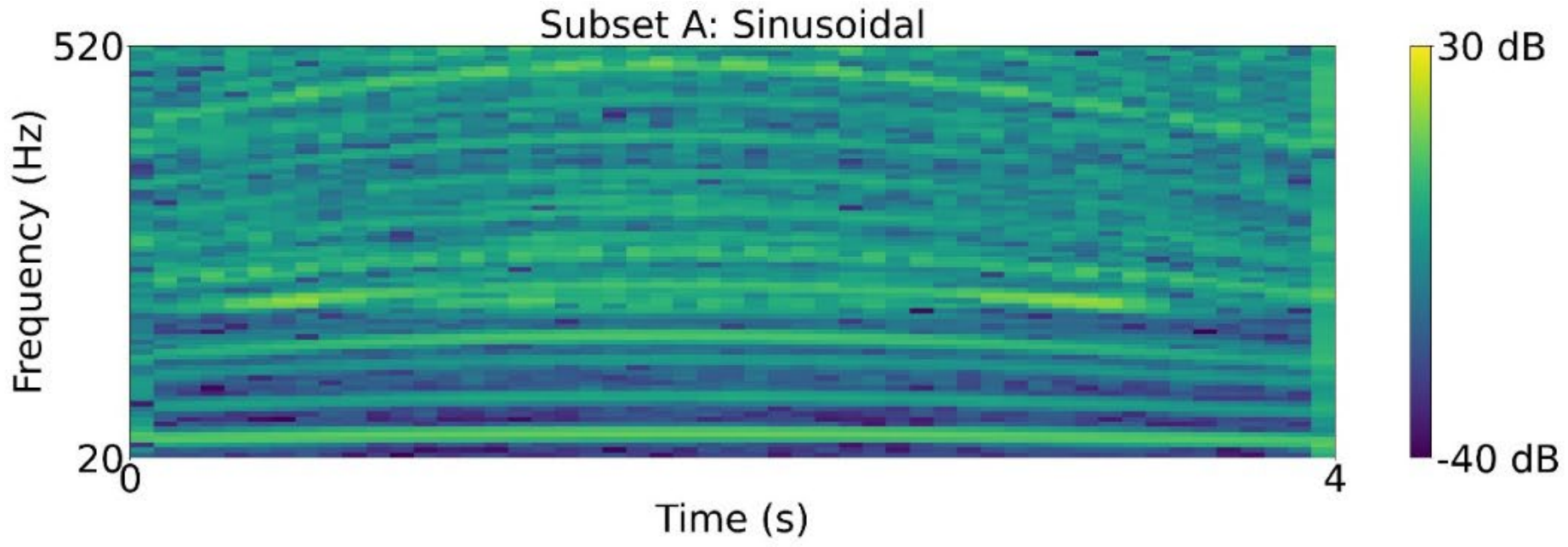

(a)

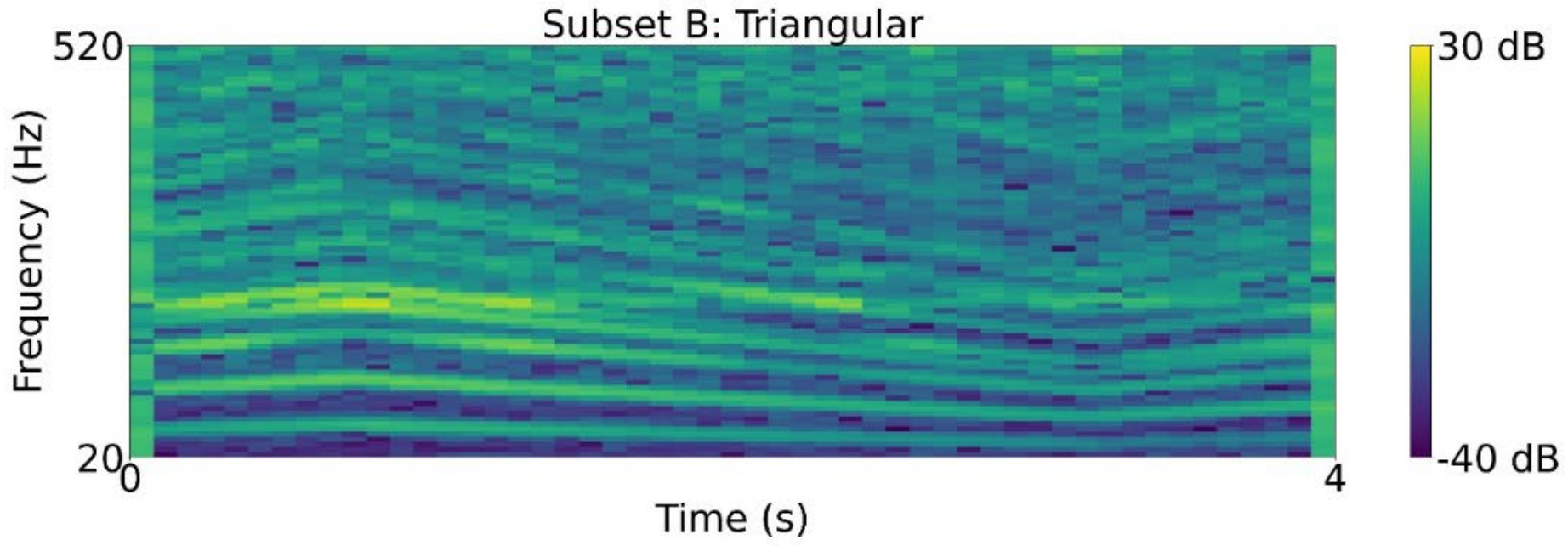

(b)

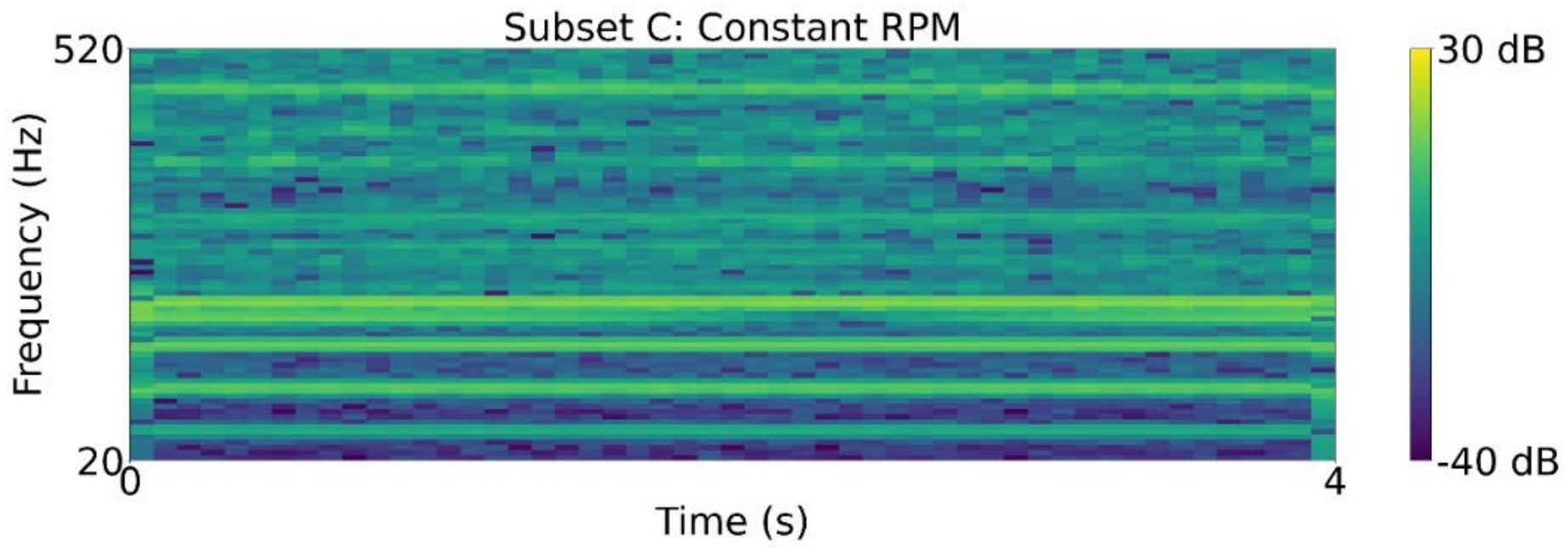

(c)

**Fig. 7.** STFT for three subsets representing different domains with varying rpm patterns. (a) Subset A with rpm variation following a sinusoidal wave, (b) Subset B with rpm variation following a triangular wave, and (c) Subset C with constant rpm.

IRF× 2 ORF × 3 misalignment × 3 unbalance) exist. The way of implementing unbalance is shown in **Fig. 6 (h)**.

To implement various operating conditions, we experimented with various rpm patterns and torque load and defined each pattern as subset. Details of the subset are elaborated in **Table 2**. In subset A, the rpm varied in a sinusoidal function pattern, with a base rpm of 3000 rpm and a modulation period of 10 seconds. Random current was applied to the hysteresis brake to generate variable torque loads. In subset B, triangular function with a base rpm of 4000 rpm and a modulation period of 5 seconds is used for rpm variation. For subset B, no torque load was applied.

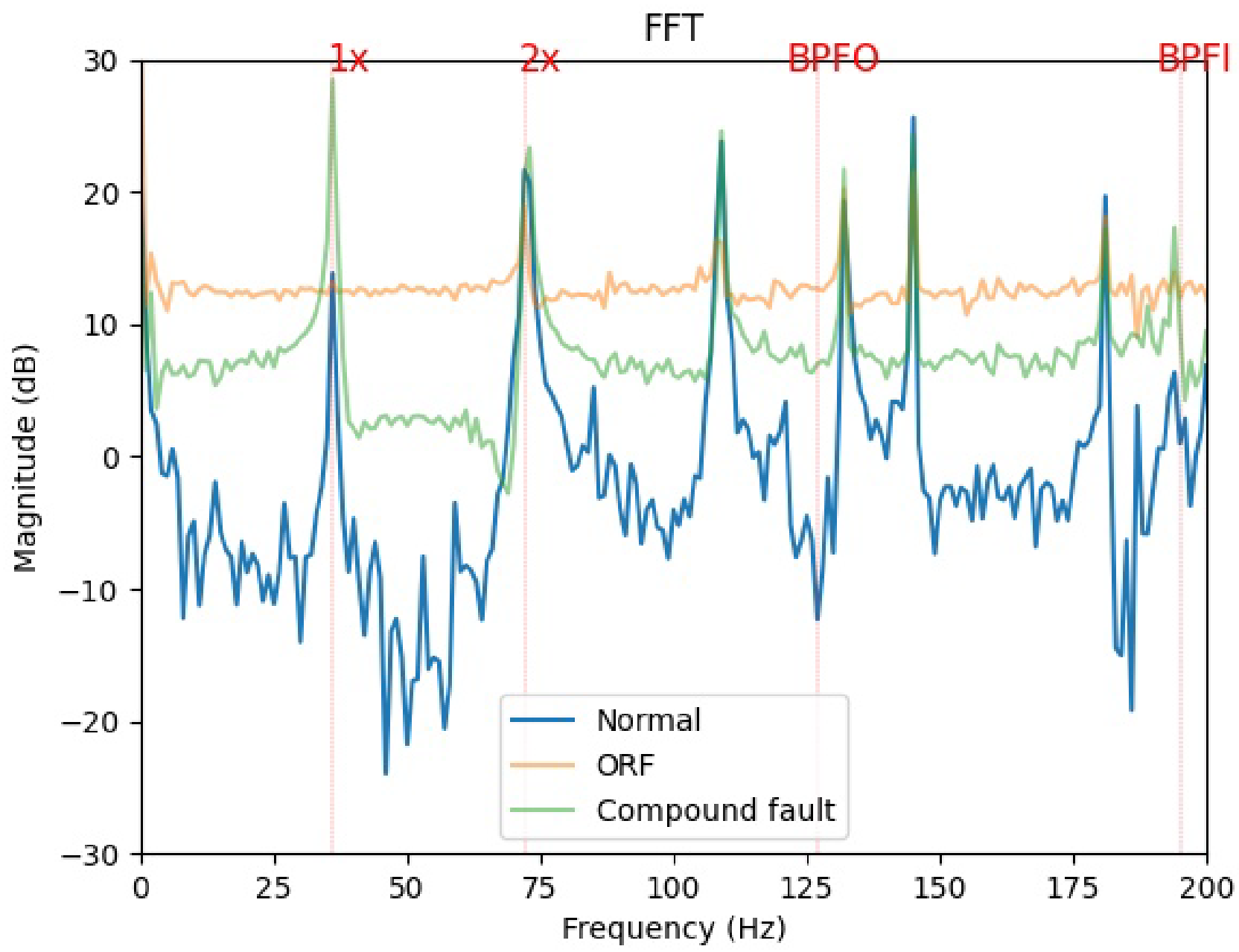

**Fig. 8.** Power spectra of vibration signals under three conditions: normal, ORF, and compound fault (IRF, ORF, misalignment, unbalance occurred simultaneously). Harmonics of shaft speed (1x, 2x), and characteristic fault frequencies (BPFO, BPFI) are indicated.

Subset C used 5 constant rpm levels corresponding to 1800 rpm, 2100 rpm, 2400 rpm, 2700 rpm, and 3000 rpm. For subsets A and B, normal data was recorded for 50 minutes, while each fault was recorded for 5 minutes. In subset C, normal data was recorded for 10 min per rpm level, whereas fault data was recorded for 1 minute per rpm level for each fault. In total, we recorded 675 minutes of vibration data for the experiment. This dataset can be downloaded from Zenodo (https://zenodo.org/records/15742618).

By conducting motor experiments under various operating conditions, particularly by varying the RPM patterns, different subsets are generated. Consequently, the time-frequency representations (TFRs) of these subsets exhibit differences as **Fig. 7**. Therefore, to ensure effective fault diagnosis across different subsets, i.e., different operating conditions, it is necessary to consider aligning features through domain adaptation.

To illustrate the spectral differences between normal and fault conditions, power spectra are analyzed to observe representative vibration signals under three classes: normal, ORF, and compound fault. As shown in Fig. 8, vibration signals from different classes exhibit distinct spectral patterns, particularly in the shaft rotation harmonics (1×, 2×), ball pass frequency outer race (BPFO), and ball pass frequency inner race (BPFI). While some spectral overlap exists, the frequency-domain features still allow effective classification across different classes.

### 5.2. *Training, validation, and test procedures*

Domain adaptation experiments were conducted across six scenarios: subset A → subset B, subset A → subset C, subset B → subset A, subset B → subset C, subset C → subset A, and subset C → subset B. These scenarios comprehensively assess model performance under different source and target domain conditions. The dataset was partitioned into training, validation, and testing sets within the source domain following an 8:1:1 ratio. While all samples in the source domain were labeled, only 10% of the target domain data were labeled, with the remaining 90% remaining unlabeled.

The training and validation process consists of two stages: pre-training in the source domain and fine-tuning in the target domain. These stages are designed to optimize model performance under domain shift conditions. For pre-training, the model was trained for 100 epochs to classify the source domain data. Similarly, for fine-tuning, the model was trained for 100 epochs to classify the target domain data. The model was optimized using the Adam optimizer [36] with a learning rate of 1e-3. The best model was selected based on the epoch with the lowest validation loss and was subsequently used for testing.

The experiments were implemented using PyTorch. All computations were conducted on a system equipped with an AMD Ryzen 7 7700 CPU and an NVIDIA GeForce RTX 4070 GPU.

The evaluation metric used for assessing model performance is the macro F1 score, computed based on 36 compound fault conditions derived from two levels of IRF, two levels of ORF, three levels of misalignment, and three levels of unbalance, including all possible combinations. The macro F1 scores across six domain adaptation scenarios, along with their average values, are presented in **Table 3** of Section 6. To ensure reliability, all results were averaged over three different random seeds.

We compare RNDR against several baseline methods, including MCC, MOC with STL, denoted as MOC (STL), TCDCN [20], MTAN [21], cross-stitch network (CS) [22], cross-connected network (CC) [23], and NDDR-CNN [24]. To further analyze the behavior of each architecture under simpler diagnostic settings, we additionally conducted experiments using a dataset consisting only of normal and single-fault conditions. The results of this single-fault scenario analysis are presented in **Table 4**. Furthermore, we performed an ablation study on the normalization component of the feature extractor. **Table 5** summarizes the comparative performance of FLN against layer normalization (LN) [33], batch normalization (BN) [31], and instance normalization (IN) [37]. For ablation, we additionally ablated by doing layer normalization, computing the mean and standard deviation along the time dimension T for each data sample, channel, and frequency step, independently. We named it time layer normalization (TLN) in this paper. For **Table 3**, **Table 4**, and **Table 5**, the best-performing method for each domain adaptation scenario is highlighted in bold.

### 5.3. *Evaluation on additional benchmarks*

We conducted an additional evaluation using a compound fault dataset containing gearbox and bearing faults [38]. We used ten classes, including severity levels of gear tooth breaks and combinations of inner and outer bearing faults. However, some fault cases (e.g., bearing faults without gear tooth breaks) are absent, resulting in some output labels missing under the MOC setup. To emulate realistic class imbalance in industrial settings, the number of samples for faulty conditions was down-sampled to one-tenth of the normal class, like the dataset obtained in this paper. We used vibration data from the accelerometer to record horizontal vibration acceleration of the motor drive end, and we sliced the raw data into 4 seconds. The data is recorded with sampling frequency 12.8 kHz. The train, validation, and test datasets are partitioned into an 8:1:1 ratio. Two domain subsets were defined based on operating conditions: one varying the rotational speed (subset G) and the other varying torque load (subset H). 50% of the target domain was labeled.

When computing STFT, we used 2048 as the window size and 1024 for hop length. Since there are three fault components (gear toot break, IRF, ORF), three branches were used for the shared trunk models [20, 21], and three separately weighted feature extractors were used for the MOC (STL), and cross-talk models [22-24].

Train, validation, and test are performed in the same manner as the dataset obtained in this paper. Table 6 summarizes the classification performance under domain adaptation scenarios.

## 6. Results

Table 3 Macro F1 Score (Architecture)

| Source → Target | MCC | MOC (STL) | TCDCN | MTAN | CS | CC | NDDR-CNN | RNDR |
|---|---|---|---|---|---|---|---|---|
| Subset A → Subset B | 0.991 | 0.990 | 0.999 | 0.988 | 0.997 | **0.999** | 0.998 | **0.999** |
| Subset A → Subset C | 0.673 | 0.797 | 0.774 | 0.763 | 0.822 | **0.860** | 0.784 | 0.850 |
| Subset B → Subset A | 0.827 | 0.766 | 0.807 | 0.817 | 0.798 | 0.823 | 0.815 | **0.831** |
| Subset B → Subset C | 0.833 | 0.860 | 0.870 | 0.859 | **0.886** | 0.866 | 0.832 | 0.884 |
| Subset C → Subset A | 0.766 | 0.783 | 0.797 | 0.786 | 0.800 | 0.791 | 0.786 | **0.833** |
| Subset C → Subset B | **1.000** | 0.995 | 0.996 | 0.998 | 0.996 | 0.998 | 0.999 | 0.996 |
| Average | 0.848 | 0.865 | 0.874 | 0.868 | 0.883 | 0.890 | 0.869 | **0.899** |

Table 4 Macro F1 Score (Architecture, Normal, and single fault only)

| Source → Target | MCC | MOC (STL) | TCDCN | MTAN | CS | CC | NDDR-CNN | RNDR |
|---|---|---|---|---|---|---|---|---|
| Subset A → Subset B | 0.938 | 0.927 | 0.940 | 0.911 | **0.953** | **0.953** | 0.943 | 0.945 |
| Subset A → Subset C | 0.907 | 0.858 | 0.878 | 0.888 | 0.882 | **0.929** | 0.651 | 0.907 |
| Subset B → Subset A | **0.772** | 0.655 | 0.737 | 0.677 | 0.689 | 0.721 | 0.770 | 0.734 |
| Subset B → Subset C | **0.977** | **0.977** | 0.957 | **0.977** | 0.869 | 0.959 | 0.965 | 0.967 |
| Subset C → Subset A | 0.724 | 0.714 | 0.794 | 0.736 | 0.649 | 0.695 | **0.811** | 0.773 |
| Subset C → Subset B | **1.000** | **1.000** | **1.000** | **1.000** | **1.000** | **1.000** | **1.000** | **1.000** |
| Average | 0.886 | 0.855 | 0.884 | 0.865 | 0.840 | 0.840 | 0.857 | **0.888** |

**Table 3** presents the fault classification performance across various domain adaptation scenarios. The results indicate that MOC consistently outperforms MCC across all settings. Specifically, the average macro F1 score of MCC is 0.848, whereas MOC (STL) achieves 0.865, demonstrating a clear improvement.

Within the MOC framework, MTL models exhibit superior performance compared to the STL approach. For instance, while MOC (STL) records an average macro F1 score of 0.865, MTL-based models such as TCDCN and MTAN achieve 0.874 and 0.868, respectively, highlighting the advantage of MTL over STL.

Table 5 Macro F1 Score (Normalization)

| Source → Target | LN | BN | IN | TLN | FLN |
|---|---|---|---|---|---|
| Subset A → Subset B | 0.782 | 0.998 | 0.995 | 0.594 | **0.999** |
| Subset A → Subset C | 0.664 | 0.816 | 0.830 | 0.069 | **0.850** |
| Subset B → Subset A | 0.539 | 0.794 | 0.777 | 0.426 | **0.831** |
| Subset B → Subset C | 0.662 | 0.854 | 0.861 | 0.074 | **0.884** |
| Subset C → Subset A | 0.548 | 0.785 | 0.815 | 0.179 | **0.833** |
| Subset C → Subset B | 0.868 | 0.985 | 0.988 | 0.516 | **0.996** |
| Average | 0.677 | 0.872 | 0.878 | 0.310 | **0.899** |

Table 6 Macro F1 Score (Architecture, Additional benchmark)

| Source → Target | MCC | MOC (STL) | TCDCN | MTAN | CS | CC | NDDR-CNN | RNDR |
|---|---|---|---|---|---|---|---|---|
| Subset G → Subset H | 0.443 | 0.403 | 0.311 | 0.311 | 0.413 | **0.496** | 0.413 | 0.467 |
| Subset H → Subset G | 0.306 | 0.230 | 0.222 | 0.276 | 0.379 | 0.329 | 0.205 | **0.422** |
| Average | 0.375 | 0.316 | 0.266 | 0.294 | 0.396 | 0.412 | 0.309 | **0.444** |

Among MTL-based architectures, models employing cross-talk architectures, such as CS and CC, outperform those based on a shared trunk, such as TCDCN and MTAN. The average macro F1 scores of CS and CC are 0.883 and 0.890, respectively, whereas TCDCN and MTAN achieve 0.874 and 0.868, confirming the superiority of cross-talk architectures.

The RNDR demonstrates the highest overall performance, outperforming all baseline models, including CC. Specifically, the RNDR achieves an average macro F1 score of 0.899, exceeding CC by 0.009. These results underscore the effectiveness of the RNDR in improving fault classification performance across different operating conditions.

To evaluate how the RNDR performs under a dataset containing only normal and single fault conditions, we conducted additional experiments as summarized in **Table 4**. This analysis varies the architecture while excluding compound faults, providing insight into each model's behavior in simpler diagnostic settings. The RNDR achieved the highest average macro-F1 score at 0.888, followed closely by MCC at 0.886 and TCDCN at 0.885. Unlike the compound fault scenario, where different fault types occur simultaneously and inter-task interactions are actively leveraged, the potential for such interaction is largely absent in single fault classification. This directly affects the effectiveness of the MOC architecture, which relies on multi-task synergy. Furthermore, while compound fault classification requires distinguishing among 36 classes, the single fault setting involves only seven, reducing the relative benefit of decomposed learning compared to unified approaches like MCC.

Ablation study was conducted to evaluate the impact of normalization methods within the feature extractor of the RNDR. As shown in **Table 5**, the FLN [25] achieves the highest average classification performance, reaching 0.899, while LN [33], TLN, BN [31], and IN [37] record 0.677, 0.310, 0.872, and 0.878, respectively.

In addition, **Table 6** shows that RNDR outperforms other architectures. Throughout this, we can say that RNDR works well not only with our dataset but also in general. MCC showed relatively higher performance than MOC (STL) and shared trunk models. Still, cross-talk models such as CS and CC perform better than MCC.

## 7. Discussion

To deepen our understanding of the architectural performance differences reported in **Table 3**, we examine how each model behaves in the compound fault classification setting. MCC, operating as a single-task model, lacks any mechanism to capture task-specific relationships, which limits its ability to distinguish between complex fault combinations. Also, an increasing number of classes hinders feature alignment and making decision boundary under domain adaptation scenario. In **Table 6**, MCC shows relatively high performance among other architectures since there were 10 classes but shows the lowest performance in **Table 3**, where there are 36 classes. MOC (STL) improves upon this by decomposing the classification into separate tasks, each with fewer output classes. However, due to its lack of inter-task communication, its capacity to model the interactions between concurrent fault types remains limited. Shared trunk models, such as TCDCN, address this by enforcing a common feature extractor across tasks, enabling indirect information sharing. Nevertheless, the absence of task-specific control often leads to feature entanglement, restricting the model's ability to preserve discriminative representations. Cross-talk architectures improve further by selectively transferring task-relevant information, enhancing positive transfer while minimizing task interference. The RNDR builds upon this paradigm by combining task-specific feature learning with CNN-based cross-talk interaction layers, yielding the highest classification performance among the models compared.

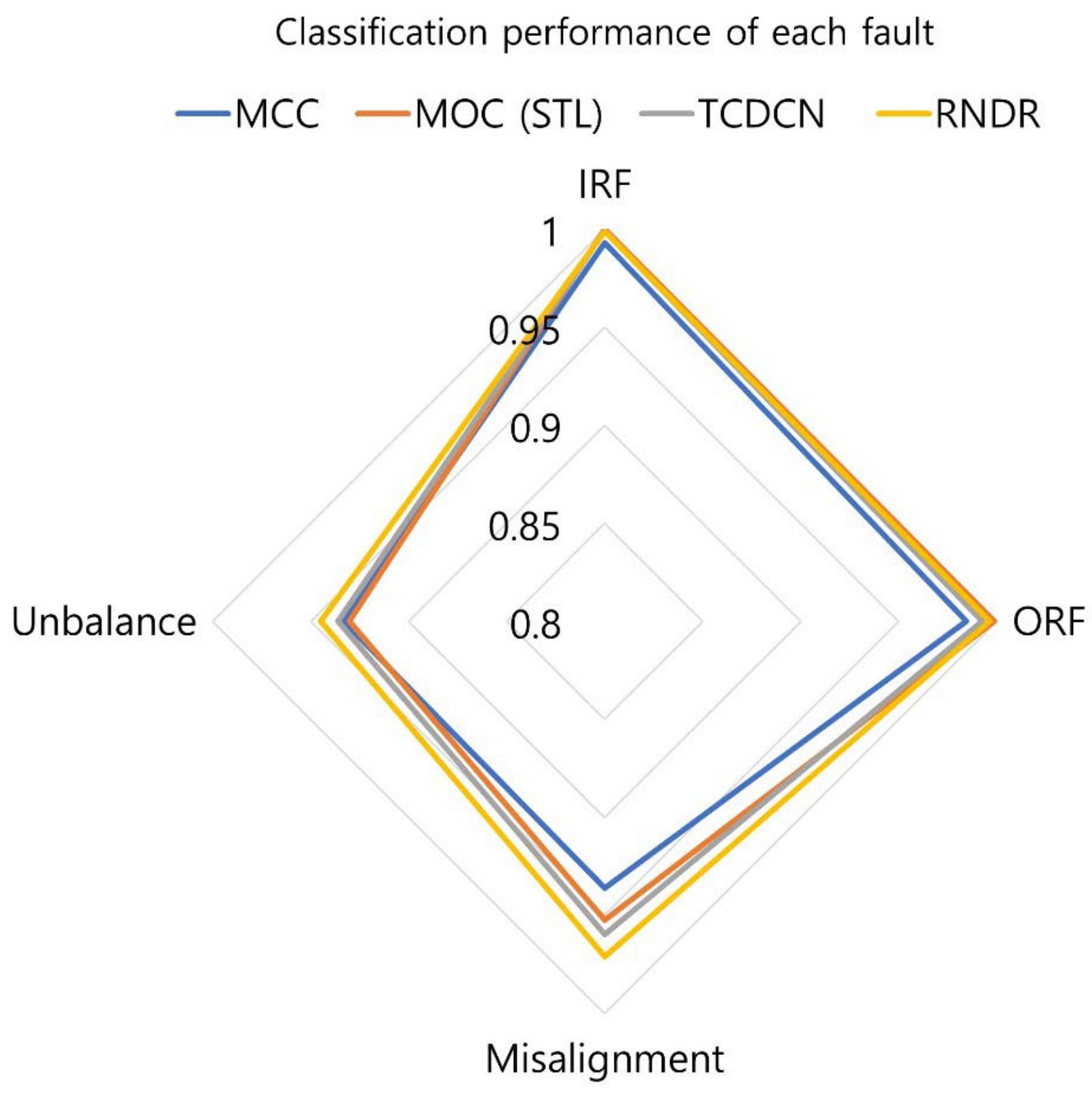

**Fig. 9.** Average macro F1 scores for IRF, ORF, misalignment, and unbalance, computed across all domain adaptation scenarios. The figure compares the classification performance of MCC, MOC (STL), TCDCN, and the RNDR, highlighting the effectiveness of different model architectures in handling compound fault diagnosis.

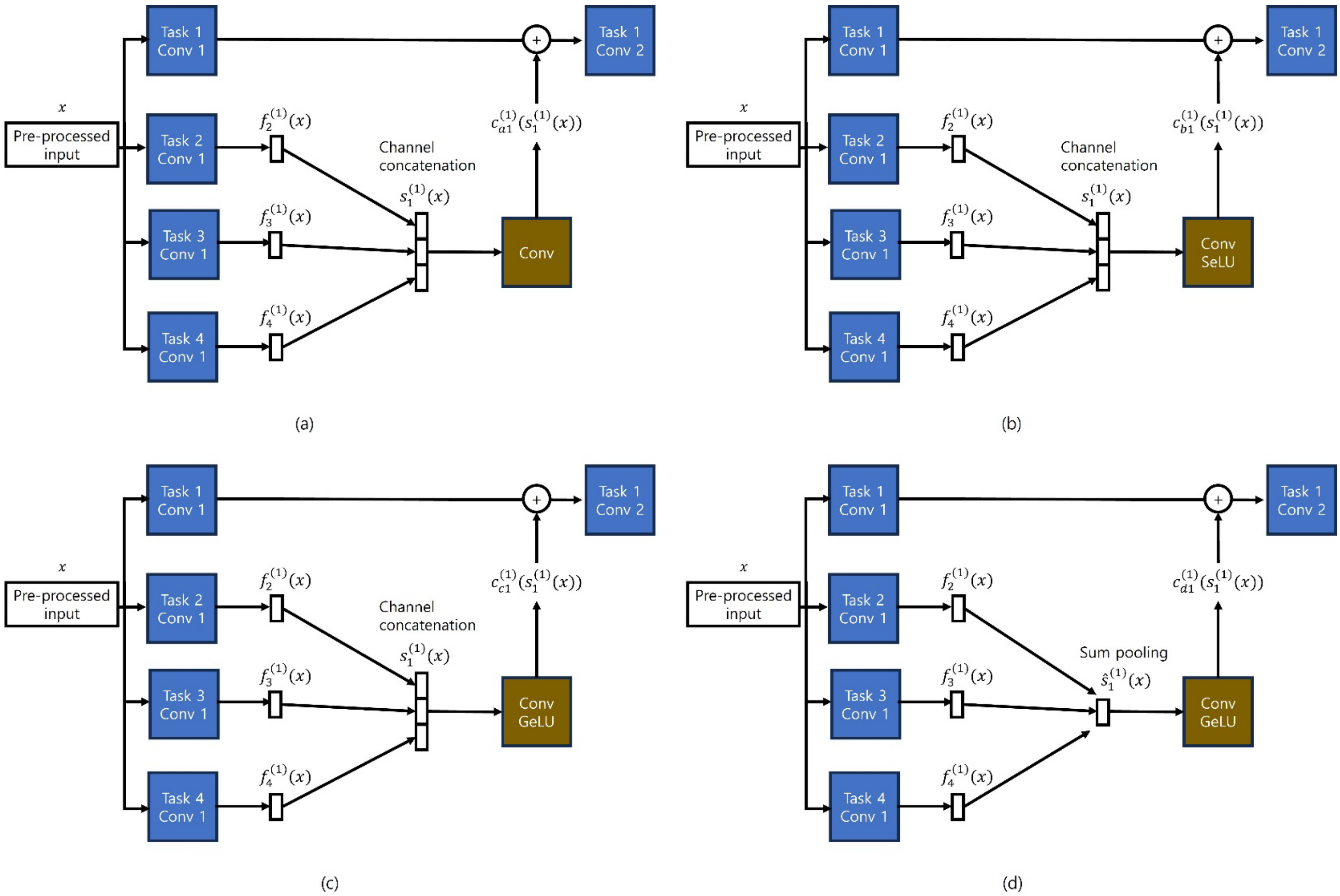

**Fig. 10.** Ablation of RNDR. (a) shows ablation model which has no ReLU activation for $c_{a1}^{(1)}$, convolution block in RNDR. (b) and (c) illustrates where activation function is changed from ReLU to SeLU and GeLU, denoting convolutional block as $c_{b1}^{(1)}$ and $c_{c1}^{(1)}$. (d) depicts the variants where channel concatenation is altered to sum pooling, named the sum pooled feature as $\hat{s}_1^{(1)}(x)$.

Beyond the experimental findings presented in the results section, an additional analysis was conducted to examine the structural merits of each architecture. While **Table 3** presents the macro F1 scores across 36 fault conditions, **Fig. 8** provides a more detailed breakdown of classification performance by showing the F1 scores for IRF, ORF, misalignment, and unbalance averaged across all domain adaptation scenarios. **Fig. 8** shows that RNDR outperforms MCC, MOC (STL), and TCDCN in fault-level classification performance.

Additionally, we ablated the proposed RNDR structure to see what made RNDR perform well. Ablated models are illustrated in Fig. 10. Fig. 10 (a) shows model that has no ReLU activation in convolutional block of RNDR block, which is depicted as $c_{a1}^{(1)}$. Fig. 10 (b) and (c) changed ReLU activation to scaled exponential linear unit (SeLU) [39] and gaussian error linear unit (GeLU) [40], respectively. Fig. 10 (d) changed channel concatenation to sum pooling, there for $\hat{s}_1^{(1)}(x) = f_2^{(1)}(x) + f_3^{(1)}(x) + f_4^{(1)}(x) \in R^{B\times C\times F\times T}$ . Since $\hat{s}_1^{(1)}(x)$ does not have larger channel dimension, convolutional block of RNDR block, $c_{d1}^{(1)}$ does not have to reduce dimension. **Table 7** shows the result for ablation study. For convenience, we notated the models displayed on Fig. 10 as ablation 1, ablation 2, ablation 3, and ablation 4 respectively.

Table 7 Macro F1 Score (Ablation study)

| | **Ablation 1** | **Ablation 2** | **Ablation 3** | **Ablation4** | **RNDR** |
|---|---|---|---|---|---|
| Average | 0.867 | 0.885 | 0.874 | 0.887 | **0.899** |

**Table 7** shows that the proposed RNDR model outperformed other ablated variants. Comparison with ablation 1 means adding ReLU activation on the convolution block of RNDR block is beneficial. Using nonlinear activation, RNDR can learn nonlinear features from interaction between fault classification tasks. The result of ablation 2 and 3 shows that using ReLU as activation function is better than using other activation functions such as SeLU and GeLU. Finally, comparing with ablation 4, we can conclude that concatenating features in the channel dimension better captures useful information than sum pooling.

Then we compared the number of parameters for every architecture we compared. The number of parameters and average macro-F1 scores across different architectures are illustrated in **Fig. 11**. MCC, being a single-task model, has the fewest parameters due to its single feature extractor and classifier. In contrast, MOC (STL) independently models each task, increasing the parameter count through separate feature extractors and task-specific classifiers. However, since it lacks inter-task communication mechanisms, its parameter size remains lower than that of cross-talk-based architectures. Among shared trunk models, TCDCN achieves relatively low model complexity by sharing a single feature extractor across all tasks. Interestingly, **Fig. 11** shows that despite this simplicity, TCDCN outperforms MOC (STL), reinforcing the notion that effective task interaction-rather than increased parameter count-is key to improving performance. MTAN, another shared trunk model, has the highest parameter count due to its task-specific attention mechanisms, yet **Fig. 11** shows that it does not yield the best classification performance. This further supports the argument that model complexity alone does not guarantee improved results. Cross-talk based architectures introduce additional layers to facilitate task-wise interaction. CS adopts a simple matrix-based transformation with minimal parameter increase, while CC, NDDR-CNN, and the RNDR employ convolutional cross-talk layers, resulting in larger model sizes. Despite this, **Fig. 11** demonstrates that the RNDR achieves the best macro-F1 score among all architectures-surpassing even more complex models such as MTAN and other cross-talk variants. These findings in **Fig. 11** collectively refute the idea that performance scales linearly with parameter size. Instead, they underscore the effectiveness of multi-task learning and the architectural benefits of selective information exchange through cross-talk layers. The RNDR, in particular, strikes an efficient balance between model complexity and inter-task communication, leading to superior classification performance in compound fault diagnosis.

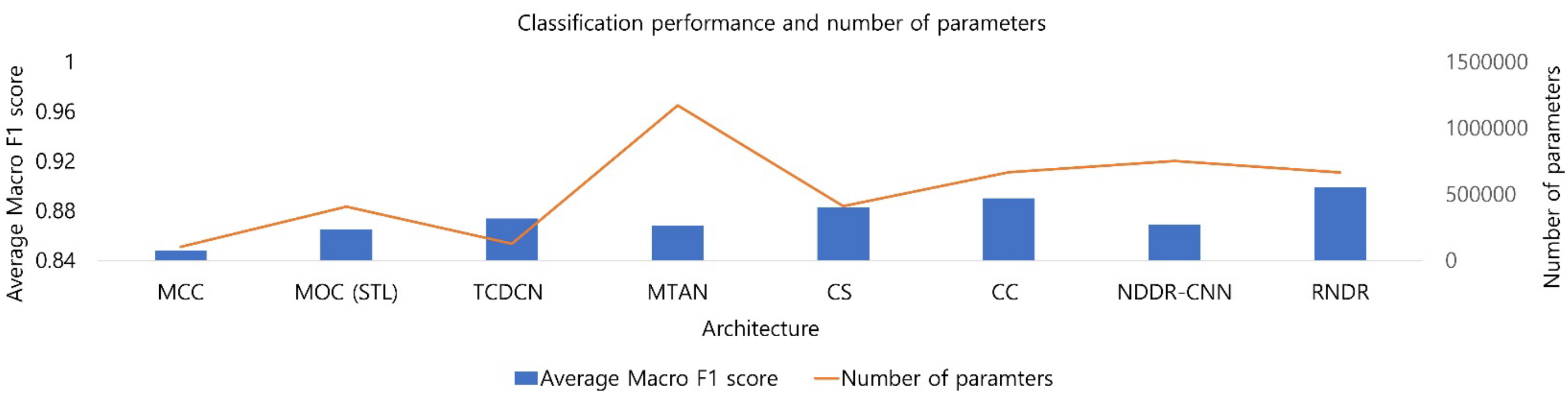

**Fig. 11.** Average macro-F1 scores (left y-axis) for compound fault classification across all domain adaptation scenarios and the number of parameters (right y-axis) for different model architectures. The left y-axis represents the classification performance in terms of macro-F1 score, while the right y-axis indicates the number of parameters for each architecture.

For all that, we increased the number of parameters of MCC, MOC (STL), and TCDCN by increasing the channel dimension and compared them with the RNDR, to ascertain the number of parameters is not only the factor of performance predominance. The result is shown in **Table 8**. For convenience, we noted model that the number of parameters is increased as size-up.

Table 8 Macro F1 Score (Comparison with the models whose number of parameters increased)

| | **MCC** | **MCC size-up** | **MOC (STL)** | **MOC (STL) size-up** | **TCDCN** | **TCDCN size-up** | **RNDR** |
|---|---|---|---|---|---|---|---|
| Average performance | 0.848 | 0.846 | 0.865 | 0.860 | 0.874 | 0.878 | **0.899** |
| Number of parameters | 103708 | 845276 | 405864 | 3328744 | 126768 | 930626 | 665066 |

**Table 8** indicates that the increase in the number of parameters does not make other models have better classification performance than RNDR. This can be strong clue for RNDR that the number of parameters is not the key to the high classification performance.

We additionally experimented with various conditions. First, we experimented the classification performance under various noise environments, since strong noise interference can exist in actual industrial environments. We add additive noise that has harmonic noise and broadband noise. Harmonic noise simulates other vibration sources that vibrate at a certain frequency, while broadband noise simulates random noise. We added noise in SNR 30 dB and 10 dB in 20~520 Hz, where we used as a frequency range for setting STFT input. Examples of signal and noise in SNR 10 dB are illustrated in **Fig. 12.** The classification performance is in **Table 9**.

Table 9 Macro F1 Score (Noise added)

| **Noise environment, SNR** | **MCC** | **MOC (STL)** | **TCDCN** | **RNDR** |
|---|---|---|---|---|
| Clean signal | 0.848 | 0.865 | 0.874 | **0.899** |
| Harmonic noise, SNR=30 dB | 0.829 | 0.839 | 0.853 | **0.867** |
| Harmonic noise, SNR=10 dB | 0.475 | 0.553 | 0.570 | **0.640** |
| Broadband noise, SNR=30 dB | 0.841 | 0.855 | 0.861 | **0.884** |
| Broadband noise, SNR=10 dB | 0.740 | 0.727 | 0.775 | **0.800** |

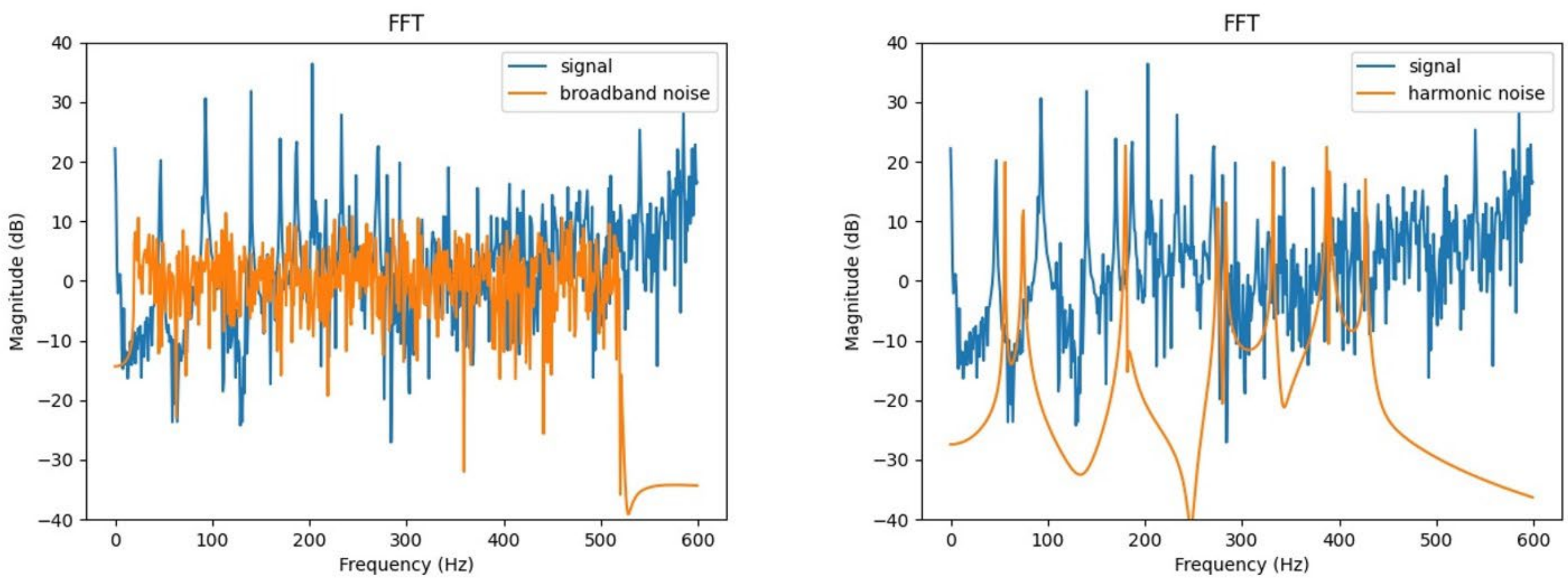

(a) (b)

**Fig. 12.** Power spectrum of signal and noise signal when SNR is 10 dB. (a) illustrates broadband noise and (b) depicts harmonic noise.

The classification performance in **Table 9** suggests that the RNDR robustly outperforms the others under different noise types and noise levels. Also, we can observe that models are more robust to broadband noise than harmonic noise. Moreover, higher levels of noise make the classification performance lower.

Then, we modified the ratio of the labeled data in the target domain dataset to 5% and 20%. The reason for this experiment is to explore the sensitivity to model performance. The result is in **Table 10**.

Table 10 Macro F1 Score (The ratio of the labeled data in the target domain modified)

| **Labeled ratio** | **MCC** | **MOC (STL)** | **TCDCN** | **RNDR** |
|---|---|---|---|---|
| 5% | 0.652 | 0.782 | 0.777 | **0.794** |
| 10% | 0.848 | 0.865 | 0.874 | **0.899** |
| 20% | 0.942 | 0.940 | 0.942 | **0.951** |

**Table 10** proves that RNDR shows the highest classification performance, even with the different labeled ratios. In addition, we can notice that the increase of labeled ratio results in higher classification performance.

Lastly, we trained and validated models with various domain adaptation losses such as correlation alignment (CORAL) [41], and maximum mean square discrepancy (MMSD) [42] to see the proposed RNDR matches well with other domain adaptation methods. The result can be seen in **Table 11**.

Table 11 Macro F1 Score (Using other domain adaptation loss)

| **Domain adaptation loss** | **MCC** | **MOC (STL)** | **TCDCN** | **RNDR** |
|---|---|---|---|---|
| MKMMD | 0.848 | 0.865 | 0.874 | **0.899** |
| CORAL | 0.832 | 0.853 | 0.870 | **0.902** |
| MMSD | 0.840 | 0.853 | 0.852 | **0.879** |

**Table 11** reveals that RNDR performs well with other domain adaptation methods. In case of CORAL, RNDR even outperformed the performance with MKMMD. However, usually, MKMMD showed the highest performance for other architectures.

## 8. Conclusion

We introduced the RNDR to improve compound fault diagnosis under domain adaptation in a partially labeled condition. Experimental results in **Table 3** demonstrated that the RNDR outperforms MCC, MOC (STL), TCDCN [20], MTAN [21], CS [22], CC [23], and NDDR-CNN [24], achieving the highest average macro F1 score of 0.899. In comparison, MCC recorded 0.848, MOC (STL) achieved 0.865, TCDCN reached 0.874, MTAN attained 0.868, CS achieved 0.883, CC reached 0.890, and NDDR-CNN attained 0.869, demonstrating a consistent performance improvement across all domain adaptation scenarios. These results confirm that MOC offers superior scalability and reduced inter-class interference compared to MCC, and that multi-task learning (MTL) more effectively utilizes shared information across tasks than single-task learning (STL). Moreover, the RNDR structure selectively enables information exchange across tasks, avoiding feature conflict typically observed in shared trunk architectures and effectively mitigating negative transfer while promoting positive transfer. Additionally, the RNDR's design is specifically suited to compound fault conditions. This was further supported by comparative results in **Table 4**, which show that the RNDR demonstrates its structural advantage more prominently in compound fault scenarios than in single-fault settings. Additionally, as shown in **Table 5**, the use of FLN [25] further contributed to domain-invariant feature extraction, outperforming LN [33], TLN, BN [31], and IN [37], which had macro F1 scores of 0.677, 0.310, 0.872, 0.878, respectively, in the normalization ablation study. In addition, through **Table 6**, we proved that the superiority of RNDR is not limited to our dataset, but also to the other dataset [38]. We further showed why the structure of RNDR is selected through **Table 7** and showed the high performance of RNDR is not due to the number of parameters by **Table 8**. At last, we showed RNDR outperforms others in noisy environments (**Table 9**), different ratios of labeled target domain (**Table 10**), and with other domain adaptation methods [41, 42] (**Table 11**).

Despite the improvements achieved by the RNDR framework, the limitation remains. As illustrated in **Fig. 11**, although the RNDR maintains a competitive parameter count relative to other cross-talk-based architectures, its use of convolutional CTL blocks inevitably increases model complexity. As a future direction, lightweight cross-talk learning mechanisms could be explored to mitigate the trade-off between performance and model size, enabling deployment in resource-constrained industrial systems. Also, classification performance has more room to improve since RNDR cannot block negative transfer perfectly. We might use pruning of CTLs or other concepts that can fully alter CTL to better mitigate negative transfer.

## Declaration of Competing Interest

The authors declare that they have no known competing financial interests or personal relationships that could have appeared to influence the work reported in this paper.

## Acknowledgments

This work was supported by the National Research Foundation of Korea (NRF) grant funded by the Korea government (MSIT) (No. RS-2024-00350917).